\documentclass{article}

\usepackage{arxiv}

\usepackage[utf8]{inputenc} 
\usepackage[T1]{fontenc}    
\usepackage[hidelinks]{hyperref}       
\usepackage{url}            
\usepackage{booktabs}       
\usepackage{amsfonts}       
\usepackage{nicefrac}       
\usepackage{microtype}      
\usepackage[authoryear,round]{natbib}
\usepackage{doi}
\usepackage{graphicx}
\usepackage[para]{threeparttable}

\title{Reproducibility and {FAIR} Principles: The Case of a Segment Polarity Network Model}

\date{April 6, 2023}	

\author{ \href{https://orcid.org/0000-0001-6507-9168}{\includegraphics[scale=0.06]{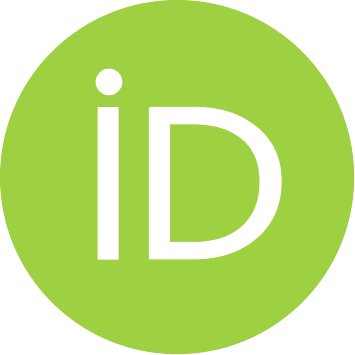}\hspace{1mm}Pedro Mendes}\\
	Center for Cell Analysis and Modeling\\
    and Department of Cell Biology\\
    University of Connecticut School of Medicine\\ Farmington, Connecticut, USA\\
	\texttt{pmendes@uchc.edu} \\
}

\renewcommand{\shorttitle}{Reproducibility and {FAIRness} of a Network Model}

\hypersetup{
pdftitle={Reproducibility and {FAIRness} of a Network Model},
pdfsubject={q-bio.NM, q-bio.QM},
pdfauthor={Pedro Mendes},
pdfkeywords={reproducibility, model reuse, computational modeling, ODE modeling, systems biology, SBML, segment polarity network},
}

\begin{document}

\maketitle

\begin{abstract}
The issue of reproducibility of computational models and the related FAIR principles (findable, accessible, interoperable, and reusable) are examined in a specific test case. I analyze a computational model of the segment polarity network in Drosophila embryos published in 2000. Despite the high number of citations to this publication, 23 years later the model is barely accessible, and consequently not interoperable. Following the text of the original publication allowed successfully encoding the model for the open source software COPASI. Subsequently saving the model in the SBML format allowed it to be \textit{reused} in other open source software packages. Submission of this SBML encoding of the model to the BioModels database enables its \textit{findability} and \textit{accessibility}. This demonstrates how the FAIR principles can be successfully enabled by using open source software, widely adopted standards, and public repositories, facilitating reproducibility and reuse of computational cell biology models that will outlive the specific software used.   
\end{abstract}

\keywords{reproducibility \and  model reuse \and computational modeling \and ODE modeling \and systems biology \and SBML \and segment polarity network}

\section{Introduction}

The year 2000 is often considered to mark the beginning of the modern systems biology era. This derives from several events that happened in that year, such as the founding of the Institute for Systems Biology, the first International Conference for Systems Biology, and the publication of various articles that are now considered ``classics''. One of those publications, by \citet{vonDassow2000}, describes a model of the Drosophila segment polarity network, where a gene regulatory network operates in each one of a series of neighboring cells, with their protein products also interacting across cells (hereafter, the ``SPN model''). The main conclusion, from a set of computer simulations sampling the SPN model's parameter space, was that it is ``remarkably'' robust as many more random combinations of parameter values  than expected give rise to the characteristic spatial gene expression pattern  required for segmentation. The inference that the network structure, rather than a narrow set of parameter values, is determinant to the phenotype has been cited as a general property of systems by more than one thousand publications to this date. Another conclusion derived from those results is that the phenotype is therefore robust against perturbation of the parameters --- and this has also frequently been assumed to be a general property of biological systems. 

An important activity in computational systems biology is the deposition of models in public repositories using standard formats like SBML \citep{Hucka2003} or CellML \citep{Hedley2001}. This allows any scientist to easily find and access those models and use them to run simulations or derive new ones using several compatible software applications. Through the last couple decades most classic models have been added to model repositories. 

Surprisingly, being described in such a highly cited publication, the SPN model is not available in any of the four major systems biology model repositories: BioModels database \citep{LeNovere2006, Sheriff2020}, the Physiome model repository \citep{Yu2011}, JWS online \citep{Olivier2004}, or the database of Virtual Cell published models \citep{Moraru2008}. To make matters worse, the software Ingeneue \citep{Meir2002,Kim2009}, used to create this model, is no longer available, not even through the Wayback Machine \citep{Archive}. Web searches revealed a SBML implementation \citep{Sethna2008} which encodes the mathematics of the model in a $4\times6$ grid of cells, but not the biochemical network.

Given the importance that the results obtained from the SPN model have had  in systems biology I felt that it should be available in a well-supported software simulator and distributed in a standard format by one of the model repositories. I therefore set to encode this model with COPASI \citep{Hoops2006,Bergmann2017} and to make sure that it was correctly implemented, use it to reproduce the simulation results of \citet{vonDassow2000}, at least partially. It has been noted that reproducing results from computational studies in general \citep{Mesirov2010,Peng2011,Stodden2016}, and  also computatational systems biology
\citep{Waltemath2016,Mendes2018, Tiwari2021}, is as hard as with laboratory experiments. This has also been the case here and the obstacles encountered are described below.

 Through a careful examination of the publications that cite \citet{vonDassow2000}, I was able to identify 15 cases where the SPN model was reused (Table \ref{table:1}). Only two actually reproduced their results \citep{Ingolia2004, Ma2006}, and another expanded the analysis to diploidy \citep{Kim_Fernandes2009}. Several authors used the SPN model to illustrate other issues, such as robustness \citep{Chaves2009,Dayarian2009,Albert2011},  ``sloppyness''  \citep{Gutenkunst2007,Daniels2008}, or new methodologies \citep{Tegner2003,Zanudo2017,Rozum2018,Marazzi2022}. Several software applications were used, such as the original Ingeneue \citep{Meir2002,Kim2009} and Little b \citep{Mallavarapu2009}, both now unavailable, and bespoke C programs that were never distributed \citep{Ingolia2004, Ma2006} --- all those results are now  difficult to reproduce.  Only the Sethna group publications \citep{Gutenkunst2007,Daniels2008} resulted in a  version of the model that is runnable in several simulators; \citet{Marazzi2022}
 re-used that model and also provided a COPASI version in their GitHub repository. 
 
This exercise identifies issues that hinder reproducibility and reuse of biomodels, and illustrates how they can be overcome with modern open science practices addressing the FAIR principles \citep{Wilkinson2016}. Reproducing it required a certain level of ``archeological'' craft to find missing parts. I hope that this also serves as a demonstration of procedures that make models usable beyond the lifetime of the software that created them. Of course, the SPN model was an important and early application of computational systems biology to developmental biology, and reproducing its results is also not irrelevant.

 \begin{threeparttable}[c]
  \centering
   \caption{Publications that reproduced or re-used  the \cite{vonDassow2000} SPN model.}
  \label{table:1}
   \begin{tabular}{ lp{7cm}ll }
    \toprule
      Reference & Description & Approach & Software \\
    \midrule
      \cite{VonDassow2002} & re-used original SPN model & ODE & Ingeneue\tnote{a} \\ 
       \cite{Albert2003} & Boolean network similar but not equal to original SPN & Boolean & unknown \\ 
        \cite{Tegner2003} & Single-cell version of original SPN, without diffusive transitions & ODE & unknown \\ 
         \cite{Ingolia2004} & re-coded original SPN model & ODE & C program\tnote{b} \\
         \cite{Ma2006} & re-coded original SPN model & ODE & C program\tnote{b} \\
         \cite{Gutenkunst2007} & re-coded original SPN model & ODE & SloppyCell\tnote{c} \\
         \cite{Daniels2008} & re-used code from \cite{Gutenkunst2007}\tnote{d} & ODE & SloppyCell\tnote{c} \\
         \cite{Chaves2009} & simplification of SPN model ODEs\tnote{e} & algebraic & N/A \\
         \cite{Dayarian2009} & simplification of SPN model ODEs\tnote{e} & algebraic & unknown \\
         \cite{Kim_Fernandes2009} & re-coded diploid version of SPN model & ODE & Mathematica\tnote{b} \\
         \cite{Mallavarapu2009} & re-coded original SPN model & ODE & Little b\tnote{a} \\
         \cite{Albert2011} & re-coded original SPN model & algebraic & MATLAB\tnote{b} \\
         \cite{Zanudo2017} & re-used original SPN model & ODE & Python\tnote{b} \\
         \cite{Rozum2018} & re-coded single-cell version of original SPN model & algebraic & Python \\
         \cite{Marazzi2022} & re-used SBML model from \cite{Daniels2008}\tnote{d} & ODE & COPASI \\
        \bottomrule
   \end{tabular}
 \begin{tablenotes}
  \item[a] Software no longer available.
  \item[b] code not publicly available.
  \item[c] Software available from https://sloppycell.sourceforge.net/ 
  \item[d] SBML version available from https://sethna.lassp.cornell.edu/Sloppy/vonDassow/model.html
   \item[e] Used a square grid of cells
 \end{tablenotes}
\end{threeparttable}

\section{Methods}

\subsection{Software}
Model simulations and parameter sampling were carried out with COPASI version 4.39 \citep[][RRID:SCR\_014260]{Hoops2006, Bergmann2017}, Virtual Cell version 7.5.0 \citep[][RRID:SCR\_007421]{Schaff1997, Moraru2008}, Tellurium version 2.2.7 \citep{Choi2018} that uses libRoadRunner version 2.3.2 \citep[][RRID:SCR\_014763]{Welsh2023}, and AMICI version 0.11.25 \citep{Froehlich2021}, which was accessed through runBioSimulations \citep[][RRID:SCR\_019110]{Shaikh2021}. The model file was constructed with python scripts using the BasiCO package that interfaces with COPASI \citep{BasiCO}. Simulations were run at the local high-performance computing cluster using the Cloud-COPASI web interface \citep{Kent2012}. Results were visualized with COPASI, with Gnuplot version 5.4.3 \citep[][RRID:SCR\_008619]{gnuplot}, or with the Python libraries Seaborn \citep{Waskom2021} (RRID:SCR\_018132) and Matplotlib \citep[][RRID:SCR\_008624]{Hunter2007}. The SBGN diagram of Figure \ref{fig:1} was created using Cell Designer version 4.4 \citep[][RRID:SCR\_007263]{Funahashi2003} and then edited with Inkscape version 1.1 (RRID:SCR\_014479).

\subsection{Model}
\begin{figure}[p!]
\begin{center}
\includegraphics[width=\textwidth]{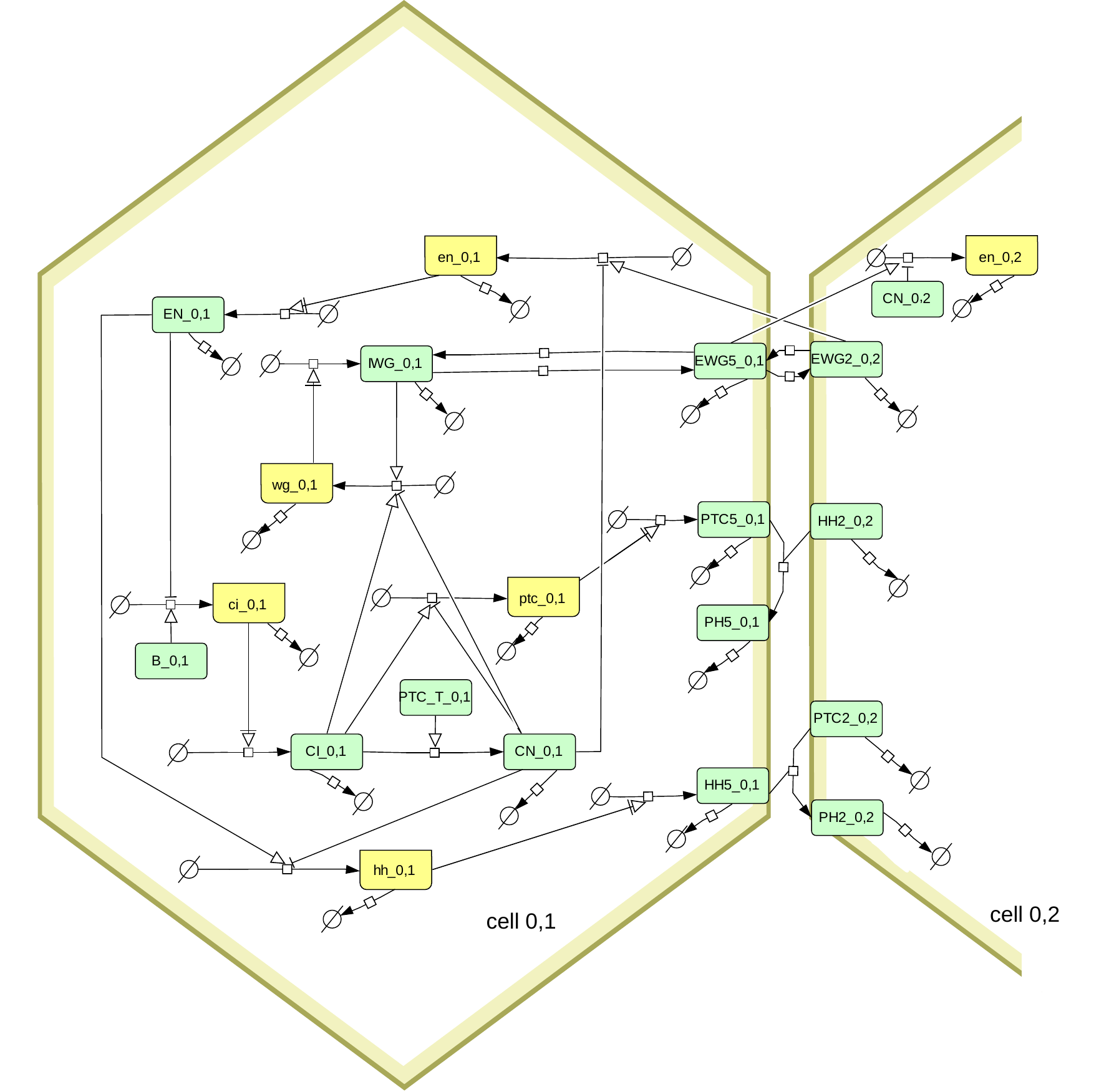}
\end{center}
\caption{ Diagram of the segment polarity network  following the SBGN standard \citep{LeNovere2009}. Boxes in light green represent proteins, boxes in yellow represent mRNA. The full model includes several hexagonal cells, this diagram shows only one (cell\_0,1) and its interactions with one of its neighbors (cell\_0,2). Note that the membrane proteins (EWG, PTC, HH, and PH) exist in six pools, one for each side of the hexagonal cell. Only the proteins in side 5 are shown on the diagram, as well as the proteins on side 2 of the neighboring cell. The membrane proteins are allowed to diffuse between sides of the hexagon, which is also not shown here ({\em eg.} EGW5\_0,1 can transfer reversibly to EGW4\_0,1 and EGW6\_0,1). The box labeled PTC\_T\_0,1 represents the sum of all PTC species (from the six sides of the membrane of cell\_0,1).}
\label{fig:1}
\end{figure}

The model used here is the segment polarity network model described by \citet{vonDassow2000}. Briefly it represents a hexagonal array of cells, where each cell can express various genes ({\it wingless}, {\it engrailed}, {\it hedgehog}, {\it cubitus interruptus}, and {\it patched}) and where their protein products interact within a cell, and across neighboring cells. Figure \ref{fig:1}  depicts the interaction network using the SBGN standard \citep{LeNovere2009}. Note that \citet{vonDassow2000} analyze two versions of this model, one having less interactions than the other. Here we only look at their full model ({\em i.e.} including the dashed arrows in the diagram of their Box 1).  Since a $1\times4$ grid of cells is enough to replicate the results \citep{vonDassow2000}, that was used here to obtain all results. 

My implementation of the model was first created for the widely used software COPASI \citep{Hoops2006,Bergmann2017} through a Python script that creates a model with arbitrary number of cells at the user's desire. A second script was created to generate the same model with only one cell, where the interacting species from neighboring cells are included as fixed concentrations. COPASI generates the full set of ODEs automatically based on the network and reaction kinetic rate laws. Unlike the SBML version from \citet{Sethna2008}, here we have the full reaction network, not just the differential equations.  A small formal difference between this version and the original SPN model, is that COPASI expresses ODEs in terms of the species amounts rather than concentrations, but since the cell volumes are not variable this makes no difference and both sets of  equations are equivalent. 

The model makes extensive use of Hill-type functions where various terms appear in the form $base^{exponent}$. This is often problematic in IEEE floating point since, for non-integer exponents, those operations are carried out based on the equivalence:

\begin{equation}
base^{exponent} = e^{exponent\times log(base)}.\label{eq:01}
\end{equation}

Therefore, calculations fail when $base$ is negative, even if infinitesimally small (generates a NaN, which in COPASI is translated to an error ``Invalid state''). Unfortunately, due to the nature of predictor-corrector integration algorithms, this can easily happen during a time course integration if one species concentration becomes very close to zero. In order to avoid this problem one can use a kind of ``guarded'' exponentiation:

\begin{equation}
base^{exponent} \simeq max(\epsilon,base)^{exponent}, \epsilon>0.\label{eq:02}
\end{equation}

Applying this protection to the model changes the rate laws. For example the rate law for transcription with inducer-repressor pair changes from the original:

\begin{equation}
 V \cdot \frac{ I \cdot \left( 1 - \frac{R^{h_2}}{k_2^{h_2} + R^{h_2}} \right)^{h_1} }  {k_1^{h_1} + I \cdot \left(1 - \frac{R^{h_2}}{k_2^{h_2} + R^{h_2}} \right)^{h_1} }  \label{eq:03}
\end{equation}

to the alternative:

\begin{equation}
V \cdot \frac{ I \cdot  max\left(\epsilon, 1 - \frac{max(\epsilon,R)^{h_2} } {k_2^{h_2} + max(\epsilon,R)^{h_2}}  \right)^{h_1} }
              { k_1^{h_1} + I \cdot  max\left(\epsilon, 1 - \frac{max(\epsilon,R)^{h_2}} {k_2^{h_2} + max(\epsilon,R)^{h_2} } \right)^{h_1} }.  \label{eq:04}
\end{equation}

The terms $k_1^{h_1}$ and $k_2^{h_2}$ are not protected by a ``guard'' because $k_1$ and $k_2$ are constants that are always positive. In the results presented here I have used $\epsilon=10^{-80}$, which reduced the incidence of simulations with NaNs from around 10\% to 0.1\%. It was never described how this problem was avoided by \citet{vonDassow2000} within the software Ingeneue. Use of these alternative rate laws was necessary for the random parameter sampling, but for specific time course simulations one can almost always use the original rate laws as described in \citet{vonDassow2000}.

Several aspects of the original SPN model were not fully described by \citet{vonDassow2000} and I have had to resort to later publications to infer what they could be. For the sake of complete transparency, here are all the details that had to be inferred from sources other than the original article:

\begin{itemize}
    \item Parameter $H_{EWG}$ does not feature in the differential equations of the Supplementary data or in \citet{VonDassow2002}, instead there the proteins $EWG$ and $IWG$ have the same half-life ($H_{IWG}$). However the parameter is clearly described as one of the 48 parameters sampled in \citet{Meir2002}, from the same group. Thus in my implementation $EWG$ has its own half-life $H_{EWG}$. 
    \item The identity of the 48 parameters that are sampled was not described unequivocally. There are in fact 53 parameters in the model (when considering 4 cells), so while 46 were obvious from their Supplementary Table S1, the other 2 could have been any of the remaining 7... Again, a Figure in \citet{Meir2002} provided the identity of the 48 parameters (which include the one mentioned in the previous bullet).
    \item The ranges for parameter samplings are provided in  Supplementary Table S1, however  it missed including the ranges for parameters $PTC_0$ and $HH_0$. \citet{Kim2009} mentions this range as $1$--$1000$ (their Table 3, parameters "max"), while an Ingeneue network file (named spg1\_01\_4cell.net), recovered from the Internet Archive \citep{Kim2010}, suggests it could be $10^{3}$--$10^{6}$. I ran simulations with both ranges, and the range $1$--$1000$ produces results closer to those reported by \citet{vonDassow2000}. 
    \item The score function used to identify parameter sets that result in the desired properties was described without sufficient detail. This scoring function is a composite of a function to identify the gene expression pattern (Eq. 15 of their Supplementary Data), and another to detect stable stripes (Eq. 16 of their Supplementary Data); the final score being the largest of these two. The text does not specify clearly what the symbols of Eq. 16 mean, particularly the $StripeScore$. Thus I only used Eq. 15 for scoring. By definition my results should identify more parameter sets than the full scoring criterion (since we are looking for scores below a threshold of 0.2).
    \item The initial conditions probed in each line of Table 1 of the original paper are not specified exactly, instead they provide ranges, such as $<20$\% value, or $20$--$60$\%, not saying whether the values used were random within that range or some actual specific values. I used $0.15$ for when they indicate $<20$\%, $0.4$ for when they specify $20$--$60$\%, and $0.9$ when they specify $60$--$100$\%. For the ``degraded'' initial condition this is even more problematic as they only provided a bar chart without axes, rather than actual values. The values I used here are specified in the Python code and in the COPASI and SBML files for the time course described below.
\end{itemize}

As described in the Interoperability section of Results, below, the model can be exported from COPASI in standard formats, particularly the systems biology markup language \citep[SBML,][]{Hucka2003,Keating2020} and the OMEX format \citep{Bergmann2014} containing a SBML file for the model and a SED-ML \citep{Waltemath2011b} file with the simulation specification.

\section{Results}

\subsection{Reproducibility}
It is rather unfortunate that the term ``reproducibility'' has itself been used with various different meanings. This confusion in terminology was discussed in detail by \citet{Goodman2016}, \citet{Plesser2018}, \citet{Milkowski2018}, and especially \citet{Barba2018}. As previously \citep{Mendes2018}, I will follow the definitions of \citet{Goodman2016}, which specifies three distinct types of reproducibility:
\begin{itemize}
\item {\it reproducibility of methods} requires one to be able to exactly reproduce the results using the same methods on the same data;
\item {\it reproducibility of results} requires one to obtain similar results in an independent study applying similar procedures;
\item {\it reproducibility of inferences} requires the same conclusions to be reached in an independent replication potentially following a different methodology.
\end{itemize}

 \begin{threeparttable}[c]
  \centering
   \caption{Frequency of solutions as a function of initial conditions.}
   \begin{tabular}{ p{6.5cm}llllll }
    \toprule
       & \multicolumn{3}{c}{\citet{vonDassow2000}} & \multicolumn{3}{c}{This work} \\
      Initial conditions & Hits & Tries & Hit rate & Hits & Tries & Hit rate \\
    \midrule
      Crisp & $1,192$ & $240,000$ & $1/201$ & $1,015$ & $239,272$ & $1/236$ \\ 
      Degraded & $149$ & $750,000$ & $1/5,000$ & $22$ & $749,988$ & $1/34,090$ \\ 
      Crisp, plus ubiquitous low-level $ci$ and $ptc$ & $110$ & $41,258$ & $1/375$ & $91$ & $41,941$ & $1/461$ \\   
      3-cell band of $ci$, $wg$ stripe on posterior margin & $69$ & $40,338$ & $1/585$ & $97$ & $41,994$ & $1/433$ \\
      3-cell band of $ptc$, $en$ stripe on anterior margin & $127$ & $36,196$ & $1/285$ & $102$ & $37,994$ & $1/372$ \\
      3-cell band of $ptc$, out-of-phase 3-cell band of $ci$ & $16$ & $226,084$ & $1/14,130$ & $168$ & $229,996$ & $1/1,369$ \\
      Close to target pattern & $464$ & $21,526$ & $1/46$ & $556$ & $21,992$ & $1/39$ \\
      \bottomrule
      \end{tabular}
 \label{table:2}
\end{threeparttable}

Because the software Ingeneue, originally used to build and simulate the SPN model, has now disappeared from circulation, reproducibility of methods can no longer be effectively carried out. In a later publication \citet{VonDassow2002} appear to have reproduced the results with the same software (see Table \ref{table:1}), however since these are the original authors, that can hardly  be seen as independent verification. Of all the works listed in Table \ref{table:1}, only \citet{Ingolia2004} and \citet{Ma2006} can be seen as independent reproductions of the original results. Unfortunately those two publications used their own C programs but did not publish them. It was work in Sethna's lab \citep{Gutenkunst2007,Daniels2008} that resulted in an electronic version of the model being created in the SBML format that is still available (see notes to Table \ref{table:1}), and which was re-used by \citet{Marazzi2022}. However this SBML implementation coded the ODEs directly without representing the reaction network,  an important limitation.

\begin{figure}[b!]
\begin{center}
\includegraphics{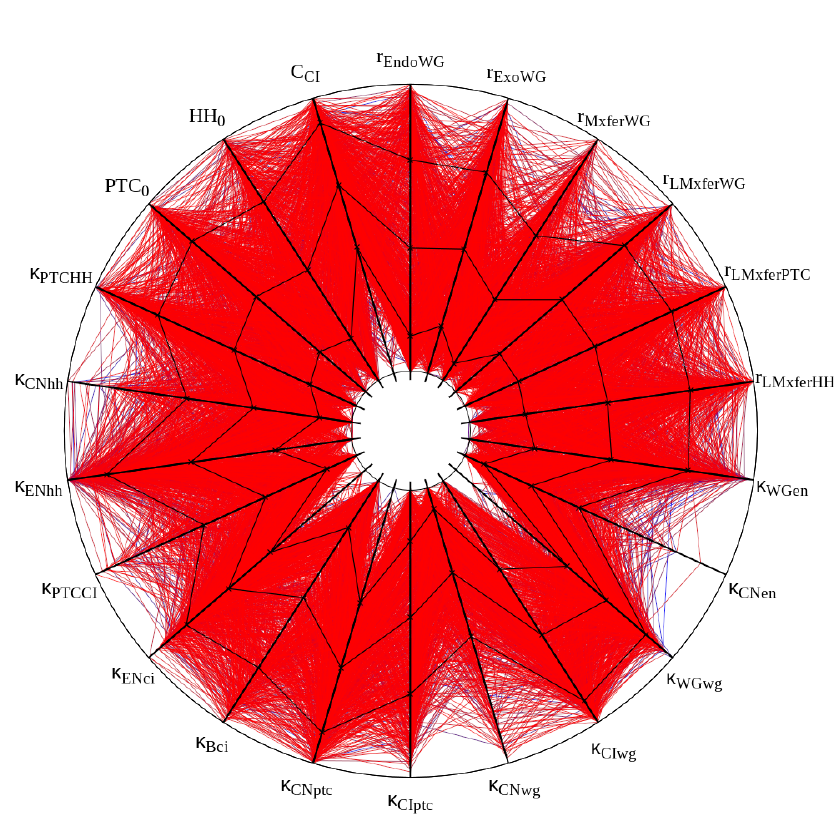}
\end{center}
\caption{Graphic representation of 'solutions’ obtained with crisp initial conditions. All 1,015 parameter sets with a  score below 0.2 are displayed. Black lines plot mean and standard deviation. Each spoke represents the log-scale range of one parameter. Half-lives and cooperativity coefficients are omitted, as in Fig. 2A of \citet{vonDassow2000}. This figure was created with the open source software Gnuplot and its source is included with the available data sets (see Data Availability Statement).}
\label{fig:2}
\end{figure}

I attempted to reproduce the results of Table 1 in \citet{vonDassow2000}, displayed in our Table \ref{table:2}. Overall these results match the original ones fairly well. There are some discrepancies in two samplings,  but these are likely due to the uncertainty on the actual initial values, as pointed out in Methods. Bear in mind that these are very small samples of a 48-dimensional parameter space and the differences may just be due to random sampling. Figure \ref{fig:2} displays the succesful parameter sets in the sampling with crisp initial conditions, corresponding to Figure 2A in \citet{vonDassow2000}. Careful comparison between the Figure and the original one reveals similar distributions. For example, in both cases $\kappa_{CNen}$ rarely takes large values. The conclusions taken by \citet{vonDassow2000} would not change if their Fig. 2A was substituted by this Figure \ref{fig:2}. Taking these results together, I propose that the current implementation of the SPN model matches the results of the original --- {\em reproducibility of results.} 

\subsection{Interoperability}
To demonstrate that this implementation of the SPN model is interoperable across different software, a specific time course was chosen to be run by several simulators (herafter named {\it timecourse1}). One of the successful parameter sets generated in the random sampling with the ``degraded'' initial condition was chosen and saved as a native COPASI file, an SBML level 3 version 1 file \citep{Keating2020}, and an OMEX file \citep{Bergmann2014}. Both the COPASI and OMEX files include the specification of the time course (end time of 1100 time units, sampled every 5 time units), though the SBML file requires that time course to be specified separately in the destination simulator.

Timecourse1 was simulated in four different software tools: COPASI, Virtual Cell \citep{Schaff1997,Moraru2008}, Tellurium \citep{Choi2018}, and AMICI \citep{Froehlich2021}. It was run locally with COPASI, Virtual Cell, and Tellurium, and through the web service runBioSimulations \citep{Shaikh2021} with AMICI. COPASI used the native file format, Tellurium used the SBML (through a small Python script ``runTellurium.py''), while Virtual Cell and AMICI used the OMEX file.

\begin{figure}[p!]
\begin{center}
\includegraphics[width=\textwidth]{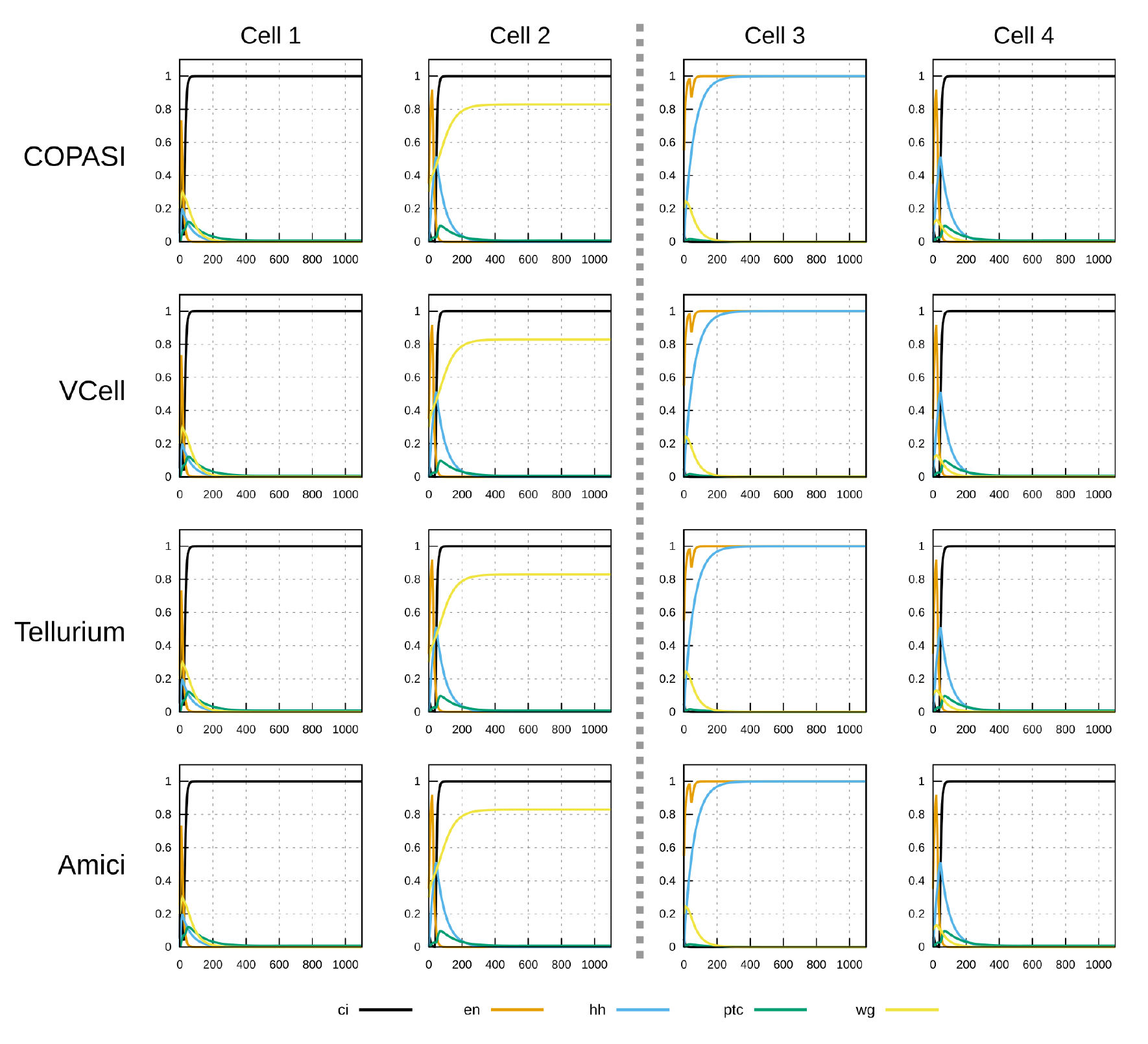}
\end{center}
\caption{Time course simulation of mRNA species in a $1\times4$ arrangement of cells using a parameter set obtained by random sampling from the ``degraded'' initial condition (see Table \ref{table:2}). Columns represent the different cells; the middle dashed line separating cell 2 and cell 3 represents a parasegmental boundary.  Displayed in each plot are the time evolution of all mRNA species in that cell. Note the formation of the expected segment polarity pattern around the parasegmental boundary, with high levels of {\it wingless} and {\it patched} in cell 2, and high levels of {\it engrailed} and {\it hedgehog} in cell 3. Each row corresponds to simulations carried out by different software. COPASI used the LSODA algorithm with absolute tolerance $10^{-13}$ and relative tolerance $10^{-8}$. Virtual Cell used a fixed step size Adams-Moulton algorithm (step size 0.1). Tellurium used CVODE non-stiff algorithm (variable step size, variable order Adams-Moulton) with absolute tolerance of $10^{-12}$ and relative tolerance of $10^{-6}$. AMICI used CVODES with absolute tolerance of $10^{-16}$ amd relative tolerance of $10^{-8}$. Results from the four simulators are visibly the same.}
\label{fig:3}
\end{figure}

\begin{figure}[p!]
\begin{center}
\includegraphics[width=\textwidth]{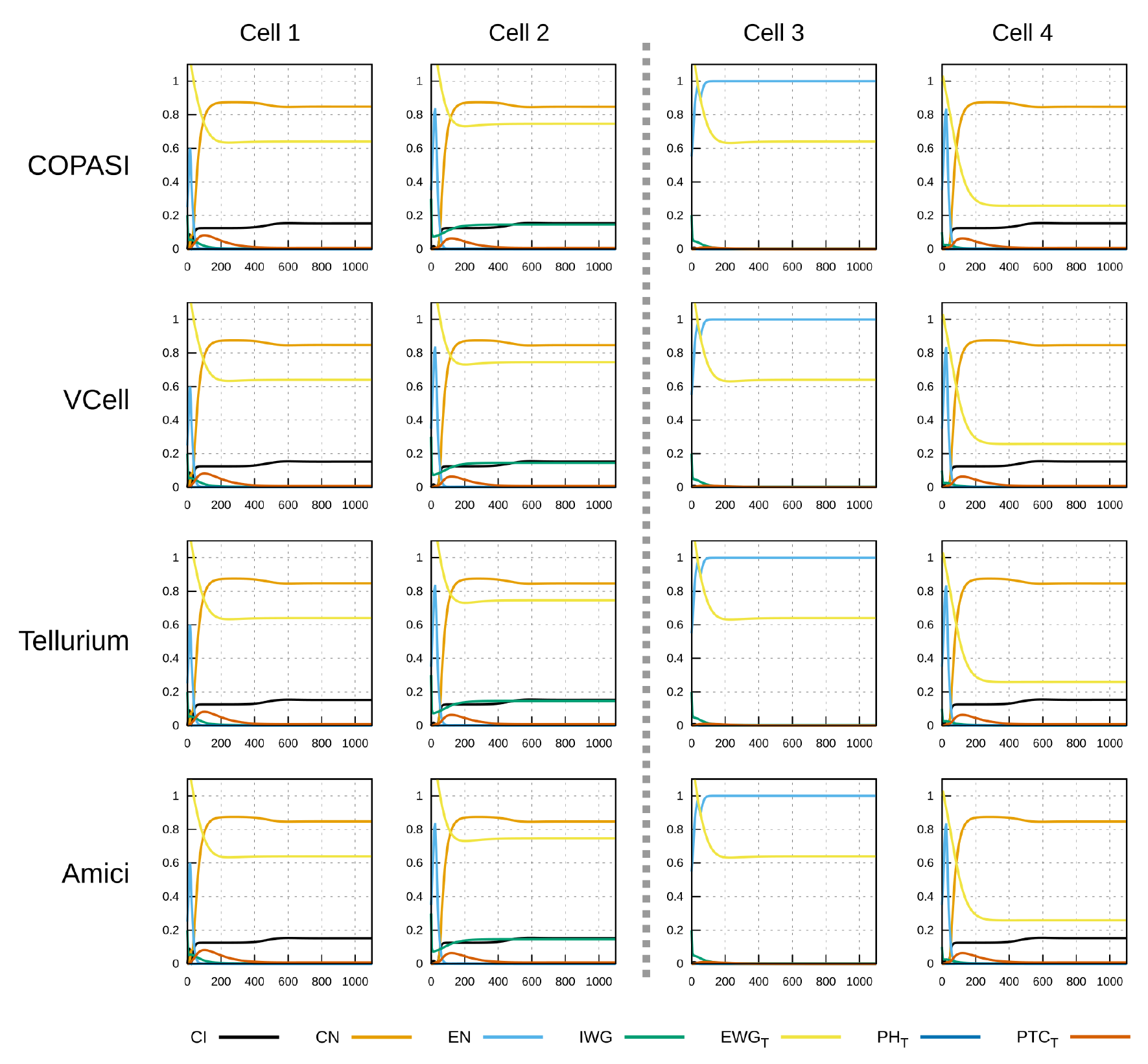}
\end{center}
\caption{Time course simulation of protein species as in Figure \ref{fig:3}. Displayed in each plot are the time evolution of some of the protein species in that cell. Species $EWG_T$ represents the total amount of $EWG$ protein (product of {\it wingless}) located in the membranes of the six neighboring cells to the one displayed; $PH_T$ is the sum of all patched–hedgehog complexes located in the six sides of that cell's membrane, and $PTC_T$ is the sum of all free patched receptor located in the six sides of that cell's membrane. Each row corresponds to simulations carried out by different software with different algorithms. As in Fig. \ref{fig:3}, there are no visible differences in the results of the  four simulators.}\label{fig:4}
\end{figure}

Figures \ref{fig:3} and \ref{fig:4} display the time course simulations obtained with four different software. There are no visible differences in the trajectories displayed confirming that these packages are all equally able to reproduce the results. Note that different ODE solvers were used by each one: COPASI used LSODA \citep{Petzold1983}, Virtual Cell used a fixed-step size Adams-Moulton method \citep{Han2002}, Tellurium used CVODE (using the Adams-Moulton variable order, variable step size method) and AMICI used CVODES, both part of the SUNDIALS suite \citep{Hindmarsh2005}.

\subsection{Findability and Accessability}

To promote findability and accessibility, the model files and associated scripts are made available through the following channels: a) a GitHub repository (https://github.com/pmendes/models/tree/main/vonDassow2000), b) a Zenodo accession DOI (doi:10.5281/zenodo.7772570), c) a submission to the Biomodels database (MODEL2304060001), and d) model files deposited in the database of public Virtual Cell models.
Note that the complete result files are only accessible through Zenodo since several files were larger than the limit at GitHub.

\subsection{Reuse}
To demonstrate how the model can be reused for different purposes, I decided to ask the question ``how often do parameter sets of the SPN model have multiple steady states?'' Earlier \citet{VonDassow2002} and especially \citet{Ingolia2004} proposed that the robustness of pattern formation in the SPN model is due to multi-stability of steady states. \citet{Ingolia2004} showed this in SPN models of a single cell (where the interacting species from the neighboring cells are kept constant). Here I investigate the answer to this question in a $1\times4$ array of cells. The strategy I used is as follows:
\begin{enumerate}
    \item generate $p$ random sets of parameter values;
    \item for each set of parameter values generate $i$ random sets of initial conditions and calculate their steady state by integration;
    \item determine how many sets of parameter values produced more than one steady state.
\end{enumerate}

COPASI can easily to carry out such study directly with the {\em Parameter scan} and {\em Steady state} tasks. The steady state task was applied here disabling the Newton method and therefore only using ODE integration to find the steady state reachable from the initial conditions (the steady state resolution was set to $10^{-4}$ and the criterion used was ``distance and time''). With the parameter scan task, 5,000 random parameter sets were sampled, using the same rules as in section 3.1 above. Then, for each parameter set, it sampled 15 random initial conditions. Since we use a model of $1\times4$ array of cells, the initial conditions are composed of 132 species concentrations that were sampled in the interval $[0,1]$. 

From the 5,000 random parameter sets generated, 3,387 had at least one steady state (the remainder are likely to contain limit cycles, but this was not investigated). Of those 3,387 parameter sets with steady states, 498 contained more than one steady state. This rate of $1/10$ parameter sets displaying multistability is not entirely surprising given the study by \cite{Ingolia2004} which highlighted the positive feedbacks contained in the SPN model. Nevertheless it is interesting to investigate if these 498 parameter sets have special characteristics versus the other 2,889 that have only one steady state. 

\begin{figure}[h!]
\begin{center}
\includegraphics{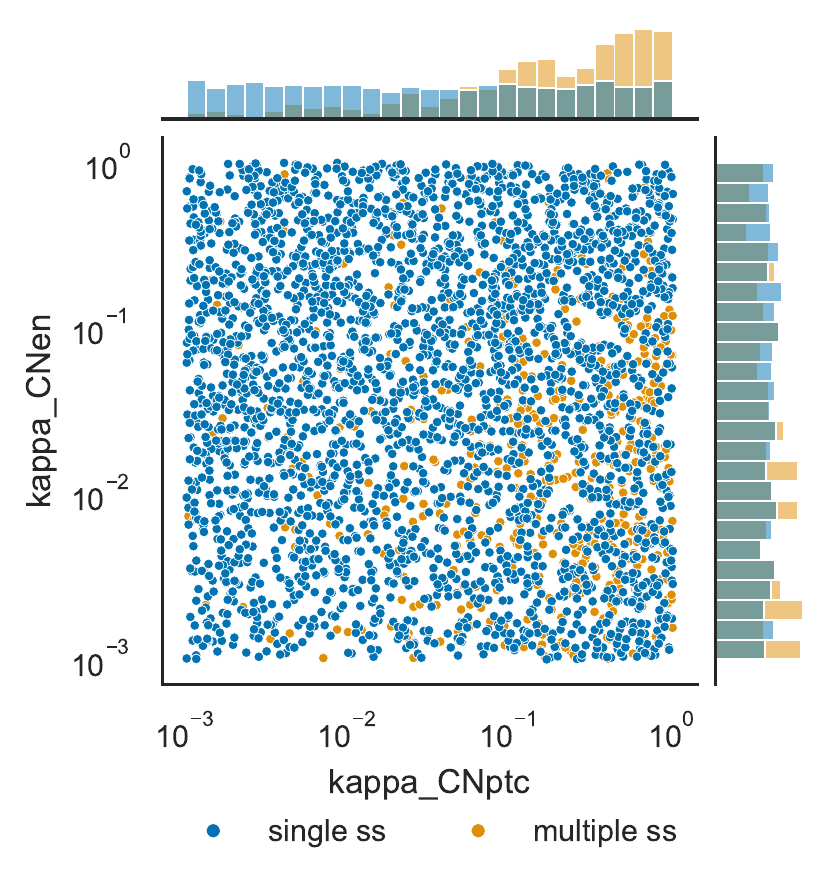}
\end{center}
\caption{Distribution of values of $\kappa_{CNptc}$ and $\kappa_{CNen}$ originating single steady states, versus those originating multiple steady states. Scatterplot of values of the two parameters and histograms of their distribution. Darker blue circles represent parameter sets for which only one steady state was identified, lighter orange circles represent parameter sets for which more than one steady state could be identified.}\label{fig:5}
\end{figure}

 The distributions of parameter values that support multiple steady states was compared with those that appear to only support a single steady state. Calculation of the relative change in the median values for each parameter in the single steady state set versus the multiple steady state set revealed that only $\kappa_{CNptc}$ shows a large difference, with a median 5-fold larger in the multiple steady state set than in the single steady state set. Three others have much lower differences: $\kappa_{CNen}$ 0.7-fold smaller, $\kappa_{CIptc}$ 0.46-fold smaller, and $HH_0$ 0.45-fold smaller. The other 44 parameters have smaller differences. Figure \ref{fig:5} depicts the distributions of values of $\kappa_{CNptc}$ and $\kappa_{CNen}$ for the two data sets. Figures \ref{fig:S1}--\ref{fig:S3} depict histograms for all of the 48 parameters. 
 There seems to be very few parameter sets that lead to multiple steady states with low values of $\kappa_{CNptc}$, while many more have high values for this parameter. This suggests that in order to achieve multiple stability the inhibition of {\it ptc} transcription by {\it CN} should be weak. This means that more cases of multistability appear when the negative feedback loop between {\it cubitus interruptus} and {\it patched} is weak. Note that there is also a positive feedback loop between these two genes (through activation of {\it ptc} transcription by {\it CI}), so reducing the strength of the negative feedback loop effectively results in a more prominent positive feedback between the two genes. \citet{Ingolia2004} found that multistability depends on the operation of the positive feedback loops. 

\section{Discussion}

It is widely recognized that there is a ``reproducibility crisis'' in science \citep{Baker2016} that includes computational science \citep{Mesirov2010, Peng2011, Stodden2016} and indeed computational modeling of biological systems \citep{Waltemath2016,Mendes2018,Tiwari2021}. I and others argue that reproducibility of results obtained from computer simulations of biological models (biomodels) could be enhanced by using open source software \citep{Ince2012,Mendes2018} that implement widely adopted standards \citep{Waltemath2016,Porubsky2021, Blinov2021}, which are part of various sets of rules proposed in the last two decades \citep{LeNovere2005, Waltemath2011, Lewis2016, Porubsky2020}. Adoption of such practices, though, will only become widespread when enforced by publishers \citep{Schnell2018,Stodden2018} and funding agencies \citep{YaleRoundtable2010}. A recent move by the US National Institutes of Health to enforce standards for data management \citep{Nih2020} is an encouraging move in that direction.

While reproducibility is a fundamental part of the scientific process \citep{Popper1959}, another important aspect is that new discoveries are almost always dependent on previous results, methodologies, and theories. To facilitate reuse of scientific data the community is increasingly adopting the so-called FAIR data principles \citep{Wilkinson2016} which promote {\em Findability}, {\em Accessibility}, {\em Interoperability}, and {\em Reuse} of data. While biomodels are usually seen as mathematics or software, they are operationally complex data objects and these principles ought to apply to them as well. Here I reproduced the reaction network, ODE model and associated simulations described in the classic systems biology paper by \citet{vonDassow2000} with the software COPASI. I then exported the model and simulation specifications in community-derived standard formats that are supported by many software applications. Finally these files were contributed to model and data repositories.  This essentially makes the model available to be manipulated by a large number of software applications, not only extant but likely future ones. Even if the standards used here will be abandoned in the future, it is most likely that converters would be developed to upgrade models to the new standards. Model and data repositories are also expected to last a long time. Thus I hope to have made this classic systems biology and development model available to a wide community, and to enable its re-use for many decades.

As in previous case studies \citep[e.g.][]{Jablonsky2011, Tiwari2021}, not all required information to reproduce the model and simulations were available in the original publication. Fortunately, there were subsequent publications by the authors and other members of their teams that hinted at the missing pieces. In some cases there is still uncertainty whether I made the correct choices, however the results obtained (Figure \ref{fig:2}) are sufficiently close to the original that these choices are at least validated to be highly plausible. This supports previous suggestions \citep{Claerbout1992,Hothorn2011,Stodden2016} that true computational reproducibility  requires availability of electronic executable versions. Unfortunately textual descriptions are almost always deficient in details, as it is only too easy to miss something. 

While the missing information in \citet{vonDassow2000} could be seen as a negative, I note that at the time the software Ingeneue was distributed together with files that allowed reproduction of the results. Additionally the model was actually described in great detail, so much that I was able to re-implement it. It is not uncommon to come across cases where even the model equations are not listed \citep[see, {\it e.g.} ][ for a survey]{Huebner2011}. However, this also highlights that publishing an electronic version alone is not guarantee that others in the future will be able to use it. In this case the software Ingeneue is no longer distributed and thus the electronic version is essentially lost (I could have tried to seek a copy from the original authors but I decided not to do so in order to test whether I could reproduce it with the information available). Publication of models in a widely used standard format is essential, as only this will assure the model to be interoperable by future software. Again, this is not a criticism of this 23 year-old publication, since at that time the relevant standards were nonexistent. 

In conclusion: we have all the tools needed to make computational systems biology models FAIR. They should be encoded in standard formats with relevant metadata and deposited in widely used repositories. Only this will assure that future researchers will be able to study and re-use these models. Any other option, such as only describing model equations, making the model available ``upon request'', or non-standard electronic encodings of the model will likely be lost within a decade or less.

\section*{Conflict of Interest Statement}
The author declares that the research was conducted in the absence of any commercial or financial relationships that could be construed as a potential conflict of interest.

\section*{Author Contributions}

PM created the concept and design of the study, run all computations, wrote the entire manuscript. PM revised, read, and approved the submitted version.

\section*{Funding}
Research reported in this publication was supported by the National Institute Of General Medical Sciences of the National Institutes of Health under Award Number R24 GM137787. The content is solely the responsibility of the author and does not necessarily represent the official views of the National Institutes of Health.

\section*{Acknowledgments}
I am grateful to Lauren Marazzi who drew my attention to the issues of findability and accessibility of this model; to Frank Bergmann who improved the BasiCO package at my request with incredible speed; to Ion Moraru and Lucian Smith for help with appropriately running Virtual Cell and Tellurium, respectively. I am also grateful to  Eran Agmon for many discussions about interoperability of biomodels.

\section*{Data Availability Statement}
The scripts and model files generated for this study can be found in the GitHub repository \url{https://github.com/pmendes/models/tree/main/vonDassow2000}. The full set of results generated from the simulations in this study can be found in the Zenodo repository with \href{https://zenodo.org/record/7772570}{DOI:10.5281/zenodo.7772570}. The model files are also available from the Biomodels database with accession number MODEL2304060001.

\bibliographystyle{agsm}
\bibliography{mendes2023}

@book{Archive,
  title={Wayback Machine},
  author={{Internet Archive}},
  year={1996},
  publisher={https://web.archive.org/},
  url={https://web.archive.org/} }

@book{NIH2020, 
  title={{NOT-OD-21-013}: Final {NIH} Policy for Data Management and Sharing}, publisher={https://grants.nih.gov/grants/guide/notice-files/NOT-OD-21-013.html}, 
  url={https://grants.nih.gov/grants/guide/notice-files/NOT-OD-21-013.html}, 
  author={{National Institutes of Health}}, 
  year={2020} }

@article{Albert2003, 
  title={The topology of the regulatory interactions predicts the expression pattern of the segment polarity genes in Drosophila melanogaster}, 
  volume={223}, 
  DOI={10.1016/s0022-5193(03)00035-3}, 
  number={1}, 
  journal={Journal of Theoretical Biology}, 
  author={Albert, Réka and Othmer, Hans G.}, 
  year={2003}, 
  pages={1–18} }

@article{Albert2011, 
  title={Computationally efficient measure of topological redundancy of biological and social networks}, 
  volume={84}, 
  DOI={10.1103/PhysRevE.84.036117}, 
  number={3 Pt 2}, 
  journal={Physical Review. E, Statistical, Nonlinear, and Soft Matter Physics}, 
  author={Albert, Réka and DasGupta, Bhaskar and Hegde, Rashmi and Sivanathan, Gowri Sangeetha and Gitter, Anthony and Gürsoy, Gamze and Paul, Pradyut and Sontag, Eduardo}, 
  year={2011}, 
  pages={036117} }

@misc{BasiCO, 
  title={basico: a simplified python interface to {COPASI}}, 
  url={https://zenodo.org/record/7665294}, 
  DOI={10.5281/zenodo.7665294}, 
  publisher={Zenodo}, 
  author={Bergmann, Frank T.}, 
  year={2023},
  howpublished={ \url{https://zenodo.org/record/7665294} }     
  }

@article{Baker2016,
   title={Is There a Reproducibility Crisis?},
   volume={533}, 
   DOI={10.1038/533452a}, 
   number={7604}, 
   journal={Nature}, 
   author={Baker, Monya}, 
   year={2016}, 
   pages={452–454} }

@article{Barba2018, 
  title={Terminologies for Reproducible Research}, 
  url={https://arxiv.org/abs/1802.03311v1}, 
  DOI={10.48550/arXiv.1802.03311}, 
  number={arXiv:1802.03311}, 
  journal={arXiv preprint},  publisher={arXiv}, 
  author={Barba, Lorena A.}, 
  year={2018} }

@article{Bergmann2014, 
  title={{COMBINE} archive and {OMEX} format: one file to share all information to reproduce a modeling project}, 
  volume={15}, 
  DOI={10.1186/s12859-014-0369-z}, 
  number={1}, 
  journal={BMC Bioinformatics}, 
  author={Bergmann, Frank T. and Adams, Richard and Moodie, Stuart and Cooper, Jonathan and Glont, Mihai and Golebiewski, Martin and Hucka, Michael and Laibe, Camille and Miller, Andrew K. and Nickerson, David P. and Olivier, Brett G. and Rodriguez, Nicolas and Sauro, Herbert M. and Scharm, Martin and Soiland-Reyes, Stian and Waltemath, Dagmar and Yvon, Florent and Le Novère, Nicolas}, 
  year={2014}, 
  pages={369} }

@article{Bergmann2017, 
  title={{COPASI} and its applications in biotechnology}, 
  volume={261}, 
  DOI={10.1016/j.jbiotec.2017.06.1200}, 
  journal={Journal of Biotechnology}, 
  author={Bergmann, Frank T. and Hoops, Stefan and Klahn, Brian and Kummer, Ursula and Mendes, Pedro and Pahle, Jürgen and Sahle, Sven}, 
  year={2017}, 
  pages={215–220} }

@article{Blinov2021, 
  title={Practical resources for enhancing the reproducibility of mechanistic modeling in systems biology}, 
  volume={27}, 
  DOI={10.1016/j.coisb.2021.06.001}, 
  journal={Current Opinion in Systems Biology}, 
  author={Blinov, Michael L. and Gennari, John H. and Karr, Jonathan R. and Moraru, Ion I. and Nickerson, David P. and Sauro, Herbert M.}, 
  year={2021}, 
  pages={100350} }

@article{Chaves2009, 
  title={Geometry and topology of parameter space: investigating measures of robustness in regulatory networks}, 
  volume={59}, 
  DOI={10.1007/s00285-008-0230-y}, 
  number={3}, 
  journal={Journal of Mathematical Biology}, 
  author={Chaves, Madalena and Sengupta, Anirvan and Sontag, Eduardo D.}, 
  year={2009}, 
  pages={315–358} }

@article{Choi2018, 
  title={Tellurium: An extensible python-based modeling environment for systems and synthetic biology}, 
  volume={171}, 
  DOI={10.1016/j.biosystems.2018.07.006}, 
  journal={Bio Systems}, 
  author={Choi, Kiri and Medley, J. Kyle and König, Matthias and Stocking, Kaylene and Smith, Lucian and Gu, Stanley and Sauro, Herbert M.}, 
  year={2018}, 
  pages={74–79} }

@inproceedings{Claerbout1992, 
  title={Electronic documents give reproducible research a new meaning}, 
  url={http://library.seg.org/doi/abs/10.1190/1.1822162}, 
  DOI={10.1190/1.1822162}, 
  booktitle={SEG Technical Program Expanded Abstracts 1992}, 
  publisher={Society of Exploration Geophysicists}, 
  author={Claerbout, Jon F. and Karrenbach, Martin}, 
  year={1992}, 
  pages={601–604} }

@article{Daniels2008, 
  title={Sloppiness, robustness, and evolvability in systems biology}, 
  volume={19}, 
  DOI={10.1016/j.copbio.2008.06.008}, 
  number={4}, 
  journal={Current Opinion in Biotechnology}, 
  author={Daniels, Bryan C. and Chen, Yan-Jiun and Sethna, James P. and Gutenkunst, Ryan N. and Myers, Christopher R.}, 
  year={2008}, 
  pages={389–395} }

@article{Dayarian2009, 
  title={Shape, size, and robustness: feasible regions in the parameter space of biochemical networks}, 
  volume={5}, 
  DOI={10.1371/journal.pcbi.1000256}, 
  number={1}, 
  journal={PLoS Computational Biology}, 
  author={Dayarian, Adel and Chaves, Madalena and Sontag, Eduardo D. and Sengupta, Anirvan M.}, 
  year={2009}, 
  pages={e1000256} }

@article{Froehlich2021,
  title={{AMICI}: High-Performance Sensitivity Analysis for Large Ordinary Differential Equation Models},
  volume={37},
  DOI={10.1093/bioinformatics/btab227},
  number={20},
  journal={Bioinformatics},
  author={Fröhlich, Fabian and Weindl, Daniel and Schälte, Yannik and Pathirana, Dilan and Paszkowski, Lukasz and Lines, Glenn Terje and Stapor, Paul and Hasenauer, Jan},
  year={2021},
  pages={3676–3677} }

@article{Funahashi2003, 
  title={CellDesigner: a process diagram editor for gene-regulatory and biochemical networks}, 
  volume={1}, 
  DOI={10.1016/S1478-5382(03)02370-9}, 
  number={5}, 
  journal={BIOSILICO}, 
  author={Funahashi, Akira and Morohashi, Mineo and Kitano, Hiroaki and Tanimura, Naoki}, 
  year={2003}, 
  pages={159–162} }

@Misc{gnuplot,
  author={ Thomas Williams and Colin Kelley and {many others} },
  title={ Gnuplot 5.4.3: an interactive plotting program },
  month={ January },
  year={ 2022 },
  howpublished={ \url{http://www.gnuplot.info/} }       
}

@article{Goodman2016, 
  title={What does research reproducibility mean?}, 
  volume={8}, 
  DOI={10.1126/scitranslmed.aaf5027}, 
  number={341}, 
  journal={Science Translational Medicine}, 
  author={Goodman, Steven N. and Fanelli, Daniele and Ioannidis, John P. A.}, 
  year={2016}, 
  pages={341ps12} }

@article{Gutenkunst2007, 
  title={Universally sloppy parameter sensitivities in systems biology models}, 
  volume={3}, 
  DOI={10.1371/journal.pcbi.0030189}, 
  number={10}, 
  journal={PLoS Computational Biology}, 
  author={Gutenkunst, Ryan N. and Waterfall, Joshua J. and Casey, Fergal P. and Brown, Kevin S. and Myers, Christopher R. and Sethna, James P.}, 
  year={2007}, 
  pages={1871–1878} }

@article{Han2002, 
  title={Solving Implicit Equations Arising from {Adams-Moulton} Methods}, 
  volume={42}, 
  DOI={10.1023/A:1021951025649}, 
  number={2}, 
  journal={BIT Numerical Mathematics}, 
  author={Han, Tian Min and Han, Yuhuan}, 
  year={2002}, 
  pages={336–350} }

@article{Hedley2001, 
  title={A short introduction to {CellML}}, 
  volume={359}, 
  DOI={10.1098/rsta.2001.0817}, 
  number={1783}, 
  journal={Philosophical Transactions of the Royal Society of London. Series A}, 
  author={Hedley, W. J. and Nelson, M. R. and Bellivant, D. P. and Nielsen, P. F.}, 
  editor={Kohl, P. and Noble, D. and Hunter, P. J.}, 
  year={2001}, 
  pages={1073–1089} }

@article{Hindmarsh2005, 
  title={SUNDIALS: Suite of nonlinear and differential/algebraic equation solvers}, 
  volume={31}, 
  DOI={10.1145/1089014.1089020}, 
  number={3}, 
  journal={ACM Transactions on Mathematical Software}, 
  author={Hindmarsh, Alan C. and Brown, Peter N. and Grant, Keith E. and Lee, Steven L. and Serban, Radu and Shumaker, Dan E. and Woodward, Carol S.}, 
  year={2005}, 
  pages={363–396} }

@article{Hoops2006, 
  title={{COPASI}---{a COmplex PAthway SImulator}}, 
  volume={22}, 
  DOI={10.1093/bioinformatics/btl485}, 
  number={24}, 
  journal={Bioinformatics}, 
  author={Hoops, Stefan and Sahle, Sven and Gauges, Ralph and Lee, Christine and Pahle, Jürgen and Simus, Natalia and Singhal, Mudita and Xu, Liang and Mendes, Pedro and Kummer, Ursula}, 
  year={2006}, 
  pages={3067–3074} }

@article{Hothorn2011, 
  title={Case studies in reproducibility}, 
  volume={12}, 
  DOI={10.1093/bib/bbq084}, 
  number={3}, 
  journal={Briefings in Bioinformatics}, 
  author={Hothorn, T. and Leisch, F.}, 
  year={2011}, 
  pages={288–300} }

@article{Huebner2011, 
  title={Applications and trends in systems biology in biochemistry}, 
  volume={278}, 
  DOI={10.1111/j.1742-4658.2011.08217.x}, 
  number={16}, 
  journal={The FEBS journal}, 
  author={Hübner, Katrin and Sahle, Sven and Kummer, Ursula}, 
  year={2011}, 
  pages={2767–2857} }

@article{Hucka2003, 
  title={The systems biology markup language ({SBML}): a medium for representation and exchange of biochemical network models}, 
  volume={19}, 
  DOI={10.1093/bioinformatics/btg015}, 
  number={4}, 
  journal={Bioinformatics}, 
  author={Hucka, M. and Finney, A. and Sauro, H. M and Bolouri, H. and Doyle, J. C and Kitano, H. and Arkin, A. P and Bornstein, B. J and Bray, D. and Cornish-Bowden, A. and Cuellar, A. A and Dronov, S. and Gilles, E. D and Ginkel, M. and Gor, V. and Goryanin, I. I. and Hedley, W. J and Hodgman, T. C and Hofmeyr, J. H and Hunter, P. J and Juty, N. S and Kasberger, J. L and Kremling, A. and Kummer, U. and Le Novere, N. and Loew, L. M and Lucio, D. and Mendes, P. and Minch, E. and Mjolsness, E. D and Nakayama, Y. and Nelson, M. R and Nielsen, P. F and Sakurada, T. and Schaff, J. C and Shapiro, B. E and Shimizu, T. S and Spence, H. D and Stelling, J. and Takahashi, K. and Tomita, M. and Wagner, J. and Wang, J.}, 
  year={2003}, 
  pages={524–531} }

@article{Hunter2007, 
  title={Matplotlib: A 2D Graphics Environment}, 
  volume={9}, 
  DOI={10.1109/MCSE.2007.55}, 
  number={3}, 
  journal={Computing in Science \& Engineering}, 
  author={Hunter, John D.}, 
  year={2007}, 
  pages={90–95} }

@article{Ince2012, 
  title={The case for open computer programs}, 
  volume={482}, 
  DOI={10.1038/nature10836}, 
  number={7386}, 
  journal={Nature}, 
  author={Ince, Darrel C. and Hatton, Leslie and Graham-Cumming, John}, 
  year={2012}, 
  pages={485–488} }

@article{Ingolia2004, 
  title={Topology and robustness in the {Drosophila} segment polarity network}, 
  volume={2}, 
  DOI={10.1371/journal.pbio.0020123}, 
  number={6}, 
  journal={PLoS Biology}, 
  author={Ingolia, Nicholas T.}, 
  year={2004}, 
  pages={e123} }

@article{Jablonsky2011, 
  title={Modeling the Calvin-Benson cycle}, 
  volume={5}, 
  DOI={10.1186/1752-0509-5-185}, 
  journal={BMC Systems Biology}, 
  author={Jablonsky, Jiri and Bauwe, Hermann and Wolkenhauer, Olaf}, 
  year={2011}, 
  pages={185} }

@article{Keating2020, 
  title={{SBML} Level 3: an extensible format for the exchange and reuse of biological models}, 
  volume={16}, 
  DOI={10.15252/msb.20199110}, 
  number={8}, 
  journal={Molecular Systems Biology}, 
  author={Keating, Sarah M. and Waltemath, Dagmar and König, Matthias and Zhang, Fengkai and Dräger, Andreas and Chaouiya, Claudine and Bergmann, Frank T. and Finney, Andrew and Gillespie, Colin S. and Helikar, Tomáš and Hoops, Stefan and Malik-Sheriff, Rahuman S. and Moodie, Stuart L. and Moraru, Ion I. and Myers, Chris J. and Naldi, Aurélien and Olivier, Brett G. and Sahle, Sven and Schaff, James C. and Smith, Lucian P. and Swat, Maciej J. and Thieffry, Denis and Watanabe, Leandro and Wilkinson, Darren J. and Blinov, Michael L. and Begley, Kimberly and Faeder, James R. and Gómez, Harold F. and Hamm, Thomas M. and Inagaki, Yuichiro and Liebermeister, Wolfram and Lister, Allyson L. and Lucio, Daniel and Mjolsness, Eric and Proctor, Carole J. and Raman, Karthik and Rodriguez, Nicolas and Shaffer, Clifford A. and Shapiro, Bruce E. and Stelling, Joerg and Swainston, Neil and Tanimura, Naoki and Wagner, John and Meier-Schellersheim, Martin and Sauro, Herbert M. and Palsson, Bernhard and Bolouri, Hamid and Kitano, Hiroaki and Funahashi, Akira and Hermjakob, Henning and Doyle, John C. and Hucka, Michael and SBML Level 3 Community members}, 
  year={2020}, 
  pages={e9110} }

@article{Kent2012, 
  title={Condor-{COPASI}: high-throughput computing for biochemical networks}, 
  volume={6}, 
  DOI={10.1186/1752-0509-6-91}, 
  journal={BMC Systems Biology}, 
  author={Kent, Edward and Hoops, Stefan and Mendes, Pedro}, 
  year={2012}, 
  pages={91} }

@article{Kim2009, 
  title={Ingeneue: a software tool to simulate and explore genetic regulatory networks}, 
  volume={500}, 
  DOI={10.1007/978-1-59745-525-1_6}, 
  journal={Methods in Molecular Biology}, 
  author={Kim, Kerry J.}, 
  year={2009}, 
  pages={169–200} }

@misc{Kim2010, 
  title={{IngeneueInMathematica}}, 
  url={https://web.archive.org/web/20100813195616/http://rusty.fhl.washington.edu/ingeneue/mathematica.html}, 
  author={Kim, Kerry J.}, 
  year={2010},
  howpublished={ \url{https://web.archive.org/web/20100813195616/http://rusty.fhl.washington.edu/ingeneue/mathematica.html} }     }

@article{Kim_Fernandes2009, 
  title={Effects of  ploidy and recombination on evolution of robustness in a model of the segment polarity network}, 
  volume={5}, 
  DOI={10.1371/journal.pcbi.1000296}, 
  number={2}, journal={PLoS Computational Biology}, 
  author={Kim, Kerry J. and Fernandes, Vilaiwan M.}, 
  year={2009}, 
  pages={e1000296} }

@article{LeNovere2005, 
  title={Minimum information requested in the annotation of biochemical models ({MIRIAM})}, 
  volume={23}, 
  DOI={10.1038/nbt1156}, 
  number={12}, 
  journal={Nature Biotechnology}, 
  author={Le Novère, N. and Finney, A. and Hucka, M. and Bhalla, U. S and Campagne, F. and Collado-Vides, J. and Crampin, E. J and Halstead, M. and Klipp, E. and Mendes, P. and Nielsen, P. and Sauro, H. and Shapiro, B. and Snoep, J. L and Spence, H. D and Wanner, B. L}, 
  year={2005}, 
  pages={1509–15} }

@article{LeNovere2006, 
  title={{BioModels} {Database}: a free, centralized database of curated, published, quantitative kinetic models of biochemical and cellular systems}, 
  volume={34}, 
  DOI={10.1093/nar/gkj092}, 
  number={Database issue}, 
  journal={Nucleic Acids Research}, 
  author={Le Novère, N. and Bornstein, B. and Broicher, A. and Courtot, M. and Donizelli, M. and Dharuri, H. and Li, L. and Sauro, H. and Schilstra, M. and Shapiro, B. and Snoep, J. L and Hucka, M.}, 
  year={2006}, 
  pages={D689-91} }

@article{LeNovere2009, 
  title={{The Systems Biology Graphical Notation}}, 
  volume={27}, 
  DOI={10.1038/nbt.1558}, 
  number={8}, 
  journal={Nature Biotechnology}, 
  author={Le Novère, Nicolas and Hucka, Michael and Mi, Huaiyu and Moodie, Stuart and Schreiber, Falk and Sorokin, Anatoly and Demir, Emek and Wegner, Katja and Aladjem, Mirit I. and Wimalaratne, Sarala M. and Bergman, Frank T. and Gauges, Ralph and Ghazal, Peter and Kawaji, Hideya and Li, Lu and Matsuoka, Yukiko and Villéger, Alice and Boyd, Sarah E. and Calzone, Laurence and Courtot, Melanie and Dogrusoz, Ugur and Freeman, Tom C. and Funahashi, Akira and Ghosh, Samik and Jouraku, Akiya and Kim, Sohyoung and Kolpakov, Fedor and Luna, Augustin and Sahle, Sven and Schmidt, Esther and Watterson, Steven and Wu, Guanming and Goryanin, Igor and Kell, Douglas B. and Sander, Chris and Sauro, Herbert and Snoep, Jacky L. and Kohn, Kurt and Kitano, Hiroaki}, 
  year={2009}, 
  pages={735–741} }

@article{Lewis2016, 
  title={Where next for the reproducibility agenda in computational biology?}, 
  volume={10}, 
  DOI={10.1186/s12918-016-0288-x}, 
  number={1}, 
  journal={BMC Systems Biology}, 
  author={Lewis, Joanna and Breeze, Charles E. and Charlesworth, Jane and Maclaren, Oliver J. and Cooper, Jonathan}, 
  year={2016}, 
  pages={52} }

@article{Ma2006, 
  title={Robustness and modular design of the {Drosophila} segment polarity network}, 
  volume={2}, 
  DOI={10.1038/msb4100111}, 
  journal={Molecular Systems Biology}, 
  author={Ma, Wenzhe and Lai, Luhua and Ouyang, Qi and Tang, Chao}, 
  year={2006}, 
  pages={70} }

@article{Mallavarapu2009, 
  title={Programming with models: modularity and abstraction provide powerful capabilities for systems biology}, 
  volume={6}, 
  DOI={10.1098/rsif.2008.0205}, 
  number={32}, 
  journal={Journal of the Royal Society, Interface}, 
  author={Mallavarapu, Aneil and Thomson, Matthew and Ullian, Benjamin and Gunawardena, Jeremy}, 
  year={2009}, 
  pages={257–270} }

@article{Marazzi2022, title={{NETISCE}: a network-based tool for cell fate reprogramming}, volume={8}, DOI={10.1038/s41540-022-00231-y}, number={1}, journal={npj Systems Biology and Applications}, author={Marazzi, Lauren and Shah, Milan and Balakrishnan, Shreedula and Patil, Ananya and Vera-Licona, Paola}, year={2022}, month={Jun}, pages={21} }

@article{Meir2002, 
  title={Ingeneue: a versatile tool for reconstituting genetic networks, with examples from the segment polarity network}, 
  volume={294}, 
  DOI={10.1002/jez.10187}, 
  number={3}, 
  journal={The Journal of Experimental Zoology}, 
  author={Meir, Eli and Munro, Edwin M. and Odell, Garrett M. and Von Dassow, George}, 
  year={2002}, 
  pages={216–251} }

@article{Mendes2018, 
  title={Reproducible Research Using Biomodels},
  volume={80}, 
  DOI={10.1007/s11538-018-0498-z}, 
  number={12}, 
  journal={Bulletin of Mathematical Biology},
  author={Mendes, Pedro}, 
  year={2018}, 
  pages={3081–3087} }

@article{Mesirov2010, 
  title={Accessible Reproducible Research}, volume={327}, 
  DOI={10.1126/science.1179653}, 
  number={5964}, 
  journal={Science}, 
  author={Mesirov, J. P.}, 
  year={2010},
  pages={415–416} }

@article{Milkowski2018, 
  title={Replicability or reproducibility? On the replication crisis in computational neuroscience and sharing only relevant detail}, 
  volume={45}, 
  DOI={10.1007/s10827-018-0702-z}, 
  number={3}, 
  journal={Journal of Computational Neuroscience}, 
  author={Miłkowski, Marcin and Hensel, Witold M. and Hohol, Mateusz}, 
  year={2018}, 
  pages={163–172} }

@article{Moraru2008, 
  title={{Virtual Cell} modelling and simulation software environment}, 
  volume={2}, 
  DOI={10.1049/iet-syb:20080102}, 
  number={5}, 
  journal={IET Systems Biology}, 
  author={Moraru, I.I. and Morgan, F. and Li, Y. and Loew, L.M. and Schaff, J.C. and Lakshminarayana, A. and Slepchenko, B.M. and Gao, F. and Blinov, M.L.}, 
  year={2008}, 
  pages={352–362} }

@article{Olivier2004, 
  title={Web-based kinetic modelling using {JWS Online}}, 
  volume={20}, 
  DOI={10.1093/bioinformatics/bth200}, 
  number={13}, 
  journal={Bioinformatics}, 
  author={Olivier, Brett G. and Snoep, Jacky L.}, 
  year={2004}, 
  pages={2143–2144} }

@article{Peng2011, 
  title={Reproducible Research in Computational Science}, 
  volume={334}, 
  DOI={10.1126/science.1213847}, 
  number={6060}, 
  journal={Science}, 
  author={Peng, R. D.}, 
  year={2011}, 
  pages={1226–1227} }

@article{Petzold1983, 
  title={Automatic selection of methods for solving stiff and nonstiff systems of ordinary differential equations.}, 
  volume={4}, 
  DOI={10.1137/0904010}, 
  number={1}, 
  journal={SIAM Journal on Scientific and Statistical Computing}, 
  author={Petzold, Linda}, 
  year={1983}, 
  pages={136–148} }

@article{Plesser2018, 
  title={Reproducibility vs. Replicability: A Brief History of a Confused Terminology}, 
  volume={11}, 
  DOI={10.3389/fninf.2017.00076}, 
  journal={Frontiers in Neuroinformatics}, 
  author={Plesser, Hans E.}, 
  year={2018}, 
  pages={76} }

@book{Popper1959, 
  address={London}, 
  title={The logic of scientific discovery}, 
  publisher={Hutchinson}, 
  author={Popper, K.R.}, 
  year={1959} }

@article{Porubsky2020, 
  title={Best Practices for Making Reproducible Biochemical Models}, 
  volume={11}, 
  DOI={10.1016/j.cels.2020.06.012}, 
  number={2}, 
  journal={Cell Systems}, 
  author={Porubsky, Veronica L. and Goldberg, Arthur P. and Rampadarath, Anand K. and Nickerson, David P. and Karr, Jonathan R. and Sauro, Herbert M.}, 
  year={2020}, 
  pages={109–120} }

@article{Porubsky2021, 
  title={Publishing reproducible dynamic kinetic models}, 
  volume={22}, 
  DOI={10.1093/bib/bbaa152}, 
  number={3}, 
  journal={Briefings in Bioinformatics}, 
  author={Porubsky, Veronica and Smith, Lucian and Sauro, Herbert M.}, 
  year={2021}, 
  pages={bbaa152} }

@article{Rozum2018, 
  title={Identifying (un)controllable dynamical behavior in complex networks}, 
  volume={14}, 
  DOI={10.1371/journal.pcbi.1006630}, 
  number={12}, 
  journal={PLoS Computational Biology}, 
  author={Rozum, Jordan C. and Albert, Réka}, 
  year={2018}, 
  pages={e1006630} }

@article{Schaff1997, 
  title={A general computational framework for modeling cellular structure and function}, 
  volume={73}, 
  DOI={10.1016/S0006-3495(97)78146-3}, 
  number={3}, 
  journal={Biophysical Journal}, 
  author={Schaff, J. and Fink, C. C. and Slepchenko, B. and Carson, J. H. and Loew, L. M.}, 
  year={1997}, 
  pages={1135–1146} }

@article{Schnell2018, 
  title={{“Reproducible”} Research in Mathematical Sciences Requires Changes in our Peer Review Culture and Modernization of our Current Publication Approach}, 
  volume={80}, 
  DOI={10.1007/s11538-018-0500-9}, 
  number={12}, 
  journal={Bulletin of Mathematical Biology}, 
  author={Schnell, Santiago}, 
  year={2018}, 
  pages={3095–3105} }

@article{Shaikh2021, 
  title={{RunBioSimulations}: an extensible web application that simulates a wide range of computational modeling frameworks, algorithms, and formats}, 
  volume={49}, 
  DOI={10.1093/nar/gkab411}, 
  number={W1}, 
  journal={Nucleic Acids Research}, 
  author={Shaikh, Bilal and Marupilla, Gnaneswara and Wilson, Mike and Blinov, Michael L. and Moraru, Ion I. and Karr, Jonathan R.}, 
  year={2021}, 
  pages={W597–W602} }

@misc{Sethna2008, 
  title={Segment polarity model}, 
  url={https://sethna.lassp.cornell.edu/Sloppy/vonDassow/model.html}, 
  author={Sethna, James P.}, 
  year={2008},
  howpublished={ \url{https://sethna.lassp.cornell.edu/Sloppy/vonDassow/model.html} }
  }

@article{Sheriff2020, 
  title={{BioModels}---15 years of sharing computational models in life science}, 
  volume={48}, 
  DOI={10.1093/nar/gkz1055}, 
  number={D1}, 
  journal={Nucleic Acids Research}, 
  author={Malik-Sheriff, Rahuman S. and Glont, Mihai and Nguyen, Tung V. N. and Tiwari, Krishna and Roberts, Matthew G. and Xavier, Ashley and Vu, Manh T. and Men, Jinghao and Maire, Matthieu and Kananathan, Sarubini and Fairbanks, Emma L. and Meyer, Johannes P. and Arankalle, Chinmay and Varusai, Thawfeek M. and Knight-Schrijver, Vincent and Li, Lu and Dueñas-Roca, Corina and Dass, Gaurhari and Keating, Sarah M. and Park, Young M. and Buso, Nicola and Rodriguez, Nicolas and Hucka, Michael and Hermjakob, Henning}, 
  year={2020}, 
  pages={D407–D415} }

@article{Stodden2016, 
  title={Enhancing reproducibility for computational methods}, 
  volume={354}, 
  DOI={10.1126/science.aah6168}, 
  number={6317}, 
  journal={Science}, 
  author={Stodden, V. and McNutt, M. and Bailey, D. H. and Deelman, E. and Gil, Y. and Hanson, B. and Heroux, M. A. and Ioannidis, J. P. A. and Taufer, M.}, 
  year={2016},
  pages={1240–1241} }

@article{Stodden2018, 
  title={An empirical analysis of journal policy effectiveness for computational reproducibility}, 
  volume={115}, 
  DOI={10.1073/pnas.1708290115}, 
  number={11}, 
  journal={Proceedings of the National Academy of Sciences USA}, 
  author={Stodden, Victoria and Seiler, Jennifer and Ma, Zhaokun}, 
  year={2018}, 
  pages={2584–2589} }

@article{Tegner2003, 
  title={Reverse engineering gene networks: integrating genetic perturbations with dynamical modeling}, 
  volume={100}, 
  DOI={10.1073/pnas.0933416100}, 
  number={10}, 
  journal={Proceedings of the National Academy of Sciences USA}, 
  author={Tegner, Jesper and Yeung, M. K. Stephen and Hasty, Jeff and Collins, James J.}, 
  year={2003}, 
  pages={5944–5949} }

@article{Tiwari2021, 
  title={Reproducibility in systems biology modelling}, 
  volume={17}, 
  DOI={10.15252/msb.20209982}, 
  number={2}, 
  journal={Molecular Systems Biology}, 
  author={Tiwari, Krishna and Kananathan, Sarubini and Roberts, Matthew G. and Meyer, Johannes P. and Sharif Shohan, Mohammad Umer and Xavier, Ashley and Maire, Matthieu and Zyoud, Ahmad and Men, Jinghao and Ng, Szeyi and Nguyen, Tung V. N. and Glont, Mihai and Hermjakob, Henning and Malik-Sheriff, Rahuman S.}, 
  year={2021}, 
  pages={e9982} }

@article{vonDassow2000, 
   author={{von} Dassow, G. and Meir, E. and Munro, E. M. and Odell, G. M.}, 
   title={The segment polarity network is a robust developmental module}, 
   volume={406},  
   number={6792}, 
   pages={188–192},
   year={2000}, 
   DOI={10.1038/35018085},  
   journal={Nature} }

@article{VonDassow2002, 
  title={Design and constraints of the {Drosophila} segment polarity module: robust spatial patterning emerges from intertwined cell state switches}, 
  volume={294}, 
  DOI={10.1002/jez.10144}, 
  number={3}, 
  journal={The Journal of Experimental Zoology}, 
  author={{von} Dassow, George and Odell, Garrett M.}, 
  year={2002}, 
  pages={179–215} }

@article{Waltemath2011, 
  title={Minimum Information About a Simulation Experiment ({MIASE})}, 
  volume={7},
  DOI={10.1371/journal.pcbi.1001122}, 
  number={4}, 
  journal={PLoS Computational Biology}, 
  author={Waltemath, Dagmar and Adams, Richard and Beard, Daniel A. and Bergmann, Frank T. and Bhalla, Upinder S. and Britten, Randall and Chelliah, Vijayalakshmi and Cooling, Michael T. and Cooper, Jonathan and Crampin, Edmund J. and Garny, Alan and Hoops, Stefan and Hucka, Michael and Hunter, Peter and Klipp, Edda and Laibe, Camille and Miller, Andrew K. and Moraru, Ion and Nickerson, David and Nielsen, Poul and Nikolski, Macha and Sahle, Sven and Sauro, Herbert M. and Schmidt, Henning and Snoep, Jacky L. and Tolle, Dominic and Wolkenhauer, Olaf and Le Novère, Nicolas}, 
  year={2011a}, 
  pages={e1001122} }

@article{Waltemath2011b, 
  title={Reproducible computational biology experiments with {SED-ML}---the Simulation Experiment Description Markup Language}, 
  volume={5}, 
  DOI={10.1186/1752-0509-5-198}, 
  journal={BMC Systems Biology}, 
  author={Waltemath, Dagmar and Adams, Richard and Bergmann, Frank T. and Hucka, Michael and Kolpakov, Fedor and Miller, Andrew K. and Moraru, Ion I. and Nickerson, David and Sahle, Sven and Snoep, Jacky L. and Le Novère, Nicolas}, 
  year={2011b}, 
  pages={198} }

@article{Waltemath2016, 
  title={How Modeling Standards, Software, and Initiatives Support Reproducibility in Systems Biology and Systems Medicine}, 
  volume={63},   DOI={10.1109/TBME.2016.2555481},number={10}, 
  journal={IEEE Transactions on Bio-Medical Engineering}, 
  author={Waltemath, Dagmar and Wolkenhauer, Olaf}, 
  year={2016}, 
  pages={1999–2006} }

@article{Waskom2021, 
  title={seaborn: statistical data visualization}, 
  volume={6}, 
  DOI={10.21105/joss.03021}, 
  number={60}, 
  journal={Journal of Open Source Software}, 
  author={Waskom, Michael L.}, 
  year={2021}, 
  pages={3021} }

@article{Welsh2023, 
  title={{libRoadRunner} 2.0: a high performance {SBML} simulation and analysis library}, 
  volume={39}, 
  DOI={10.1093/bioinformatics/btac770}, 
  number={1}, 
  journal={Bioinformatics}, 
  author={Welsh, Ciaran and Xu, Jin and Smith, Lucian and König, Matthias and Choi, Kiri and Sauro, Herbert M.}, 
  year={2023}, 
  pages={btac770} }

@article{Wilkinson2016, 
  title={The {FAIR} Guiding Principles for scientific data management and stewardship}, 
  volume={3}, 
  DOI={10.1038/sdata.2016.18}, 
  journal={Scientific Data}, 
  author={Wilkinson, Mark D. and Dumontier, Michel and Aalbersberg, I. Jsbrand Jan and Appleton, Gabrielle and Axton, Myles and Baak, Arie and Blomberg, Niklas and Boiten, Jan-Willem and da Silva Santos, Luiz Bonino and Bourne, Philip E. and Bouwman, Jildau and Brookes, Anthony J. and Clark, Tim and Crosas, Mercè and Dillo, Ingrid and Dumon, Olivier and Edmunds, Scott and Evelo, Chris T. and Finkers, Richard and Gonzalez-Beltran, Alejandra and Gray, Alasdair J. G. and Groth, Paul and Goble, Carole and Grethe, Jeffrey S. and Heringa, Jaap and ’t Hoen, Peter A. C. and Hooft, Rob and Kuhn, Tobias and Kok, Ruben and Kok, Joost and Lusher, Scott J. and Martone, Maryann E. and Mons, Albert and Packer, Abel L. and Persson, Bengt and Rocca-Serra, Philippe and Roos, Marco and van Schaik, Rene and Sansone, Susanna-Assunta and Schultes, Erik and Sengstag, Thierry and Slater, Ted and Strawn, George and Swertz, Morris A. and Thompson, Mark and van der Lei, Johan and van Mulligen, Erik and Velterop, Jan and Waagmeester, Andra and Wittenburg, Peter and Wolstencroft, Katherine and Zhao, Jun and Mons, Barend}, 
  year={2016}, 
  pages={160018} }

@article{YaleRoundtable2010, 
  title={Reproducible Research}, 
  volume={12}, 
  DOI={10.1109/MCSE.2010.113}, 
  number={5}, 
  journal={Computing in Science \& Engineering}, 
  author={{Yale Law School Roundtable Participants}}, 
  year={2010}, 
  pages={8–13} }

@article{Yu2011, 
  title={The Physiome Model Repository 2}, 
  volume={27}, 
  DOI={10.1093/bioinformatics/btq723}, 
  number={5}, 
  journal={Bioinformatics}, 
  author={Yu, Tommy and Lloyd, Catherine M. and Nickerson, David P. and Cooling, Michael T. and Miller, Andrew K. and Garny, Alan and Terkildsen, Jonna R. and Lawson, James and Britten, Randall D. and Hunter, Peter J. and Nielsen, Poul M. F.}, 
  year={2011}, 
  pages={743–744} }

@article{Zanudo2017, 
  title={Structure-based control of complex networks with nonlinear dynamics}, 
  volume={114}, 
  DOI={10.1073/pnas.1617387114}, 
  number={28}, 
  journal={Proceedings of the National Academy USA}, 
  author={Zañudo, Jorge Gomez Tejeda and Yang, Gang and Albert, Réka}, 
  year={2017}, 
  pages={7234–7239} }

\newpage

\section*{Supplemental Figures}
\renewcommand{\thefigure}{S\arabic{figure}}
\setcounter{figure}{0} 
\renewcommand*{\theHfigure}{notag.\thefigure}

\begin{figure}[h]
\begin{center}
\includegraphics[width=\textwidth,height=\textheight,keepaspectratio]{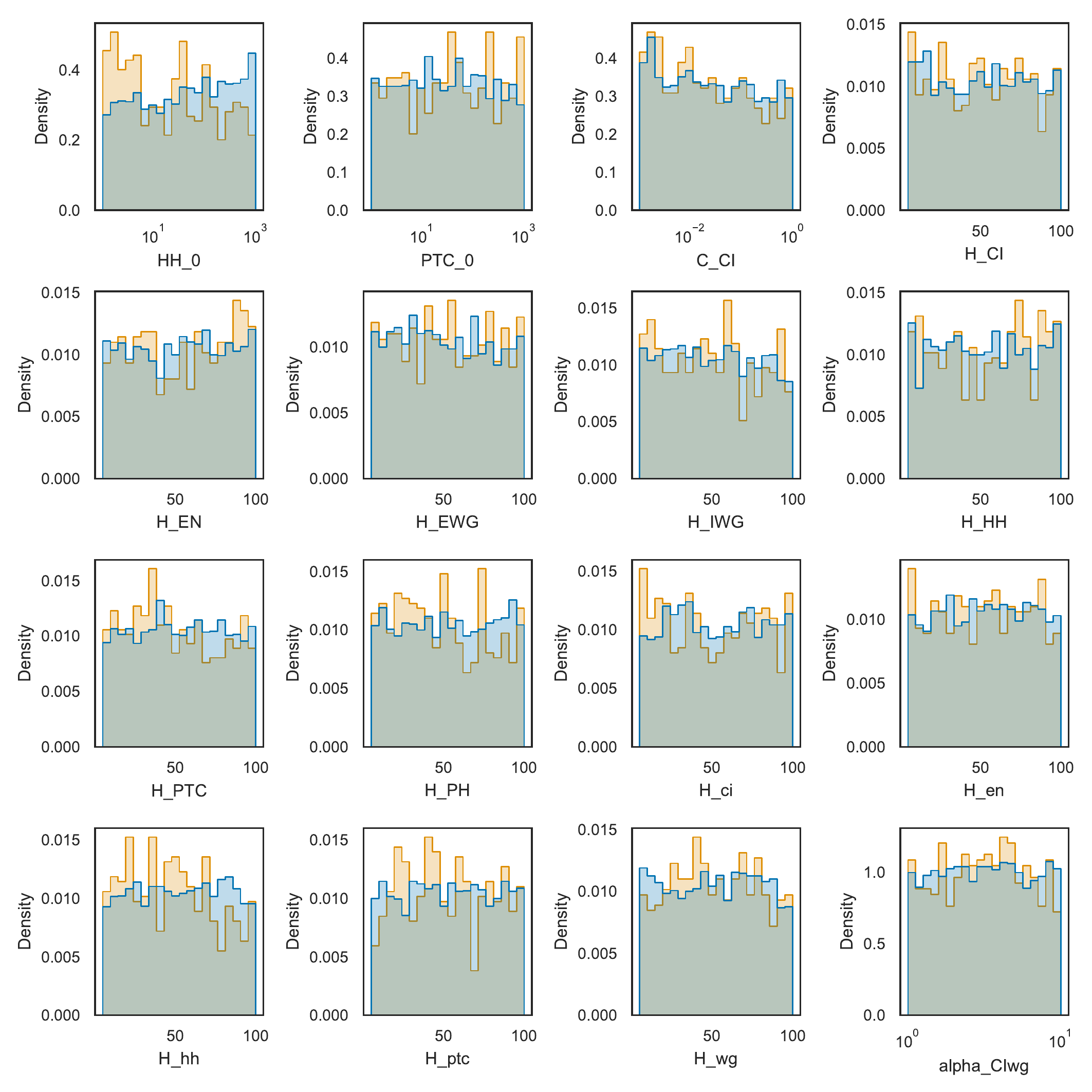}
\end{center}
\caption{ Distributions of parameter values that result in a single steady state (blue) and multiple steady states (orange).}
\label{fig:S1}
\end{figure}

\begin{figure}[h]
\begin{center}
\includegraphics[width=\textwidth,height=\textheight,keepaspectratio]{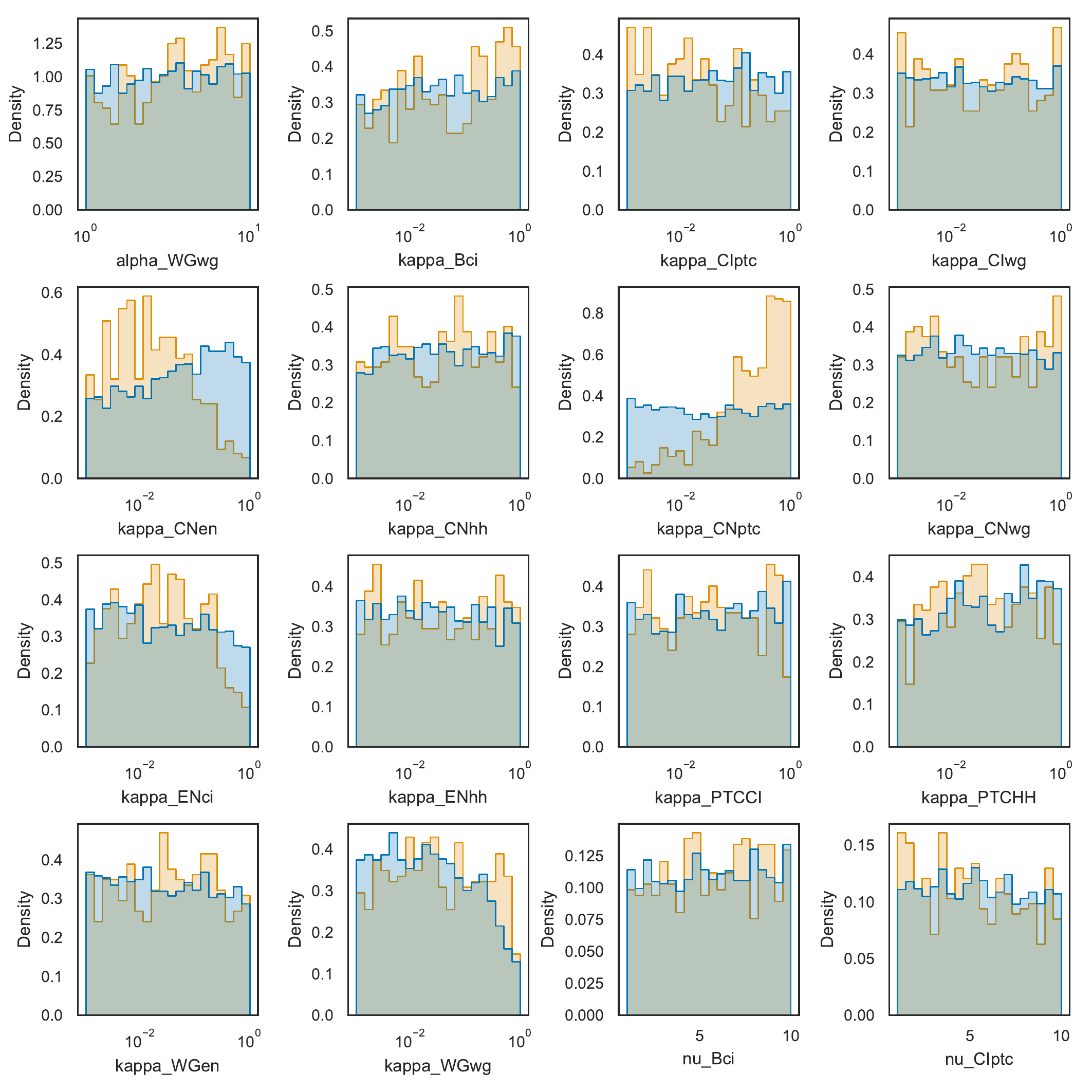}
\end{center}
\caption{ Distributions of parameter values that result in a single steady state (blue) and multiple steady states (orange). }
\label{fig:S2}
\end{figure}

\begin{figure}[h]
\begin{center}
\includegraphics[width=\textwidth,height=\textheight,keepaspectratio]{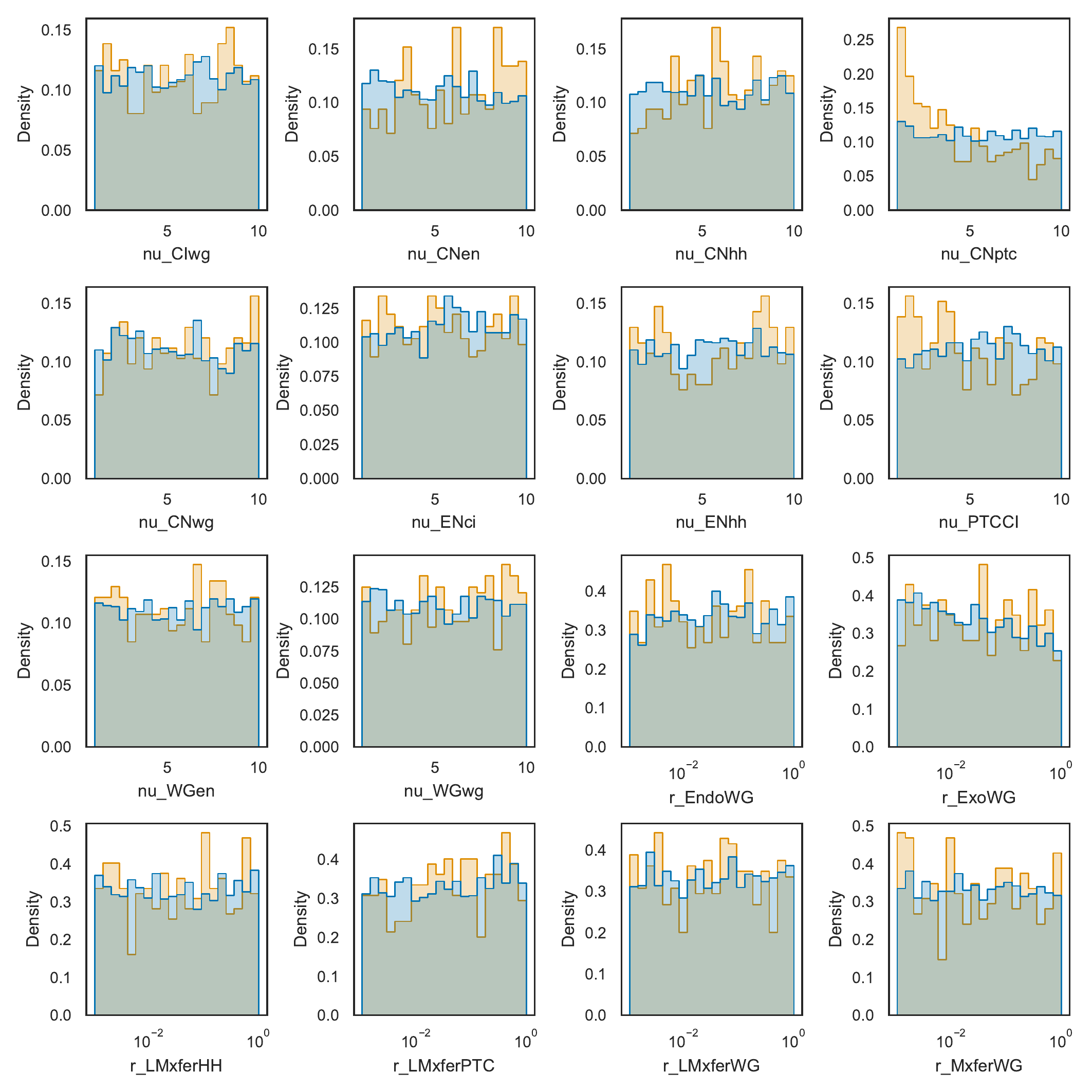}
\end{center}
\caption{ Distributions of parameter values that result in a single steady state (blue) and multiple steady states (orange).}
\label{fig:S3}
\end{figure}

\end{document}